# MAGIC: Manifold and Graph Integrative Convolutional Network for Low-Dose CT Reconstruction

Wenjun Xia[1], Zexin Lu[1], Yongqiang Huang[1], Zuoqiang Shi[2], Yan Liu[3], Hu Chen[1], Yang Chen[4], Jiliu Zhou[1], and Yi Zhang[1,*]

1 College of Computer Science, Sichuan University, Chengdu 610065, China
2 Department of Mathematical Sciences, Tsinghua University, Beijing 100084, China
3 School of Electrical Engineering Information, Sichuan University, Chengdu 610065, China
4 Laboratory of Image Science and Technology, Southeast University, Nanjing 210096, China

**Abstract:** Low-dose computed tomography (LDCT) scans, which can effectively alleviate the radiation problem, will degrade the imaging quality. In this paper, we propose a novel LDCT reconstruction network that unrolls the iterative scheme and performs in both image and manifold spaces. Because patch manifolds of medical images have low-dimensional structures, we can build graphs from the manifolds. Then, we simultaneously leverage the spatial convolution to extract the local pixel-level features from the images and incorporate the graph convolution to analyze the nonlocal topological features in manifold space. The experiments show that our proposed method outperforms both the quantitative and qualitative aspects of state-of-the-art methods. In addition, aided by a projection loss component, our proposed method also demonstrates superior performance for semi-supervised learning. The network can remove most noise while maintaining the details of only 10% (40 slices) of the training data labeled.

## 1. Introduction

Low-dose X-ray computed tomography (LDCT) can effectively reduce the risk of radiation exposure and thus plays an important role in radiology. The X-ray flux received by the patients can be reduced by switching the voltage or current of the X-ray tube. However, a lower-dose scan will degrade the signal-to-noise ratio (SNR) of the reconstructed images and compromise the diagnosis accuracy. It is very difficult to meet the diagnostic demands with LDCT images reconstructed via the classic analytical method, i.e., filtered back-projection (FBP) [1]. To balance the radiation dose and imaging quality, a number of algorithms have been developed for LDCT reconstruction. With the very recent technological innovations, these algorithms can generally be divided into two categories: 1) regularization-based methods and 2) learning-based methods.

The regularization-based methods formulate the prior knowledge into a reconstruction model. Appropriate prior information, which efficiently characterizes the target image, can maintain the critical details of the reconstructed result while eliminating unexpected noise and artifacts. At first, L2 norm regularizations, with good denoising abilities, were widely studied. In a typical investigation, the penalized weighted least squares (PWLS) algorithm was proposed by Fessler [2]. Wang *et al*. introduced the quadratic smoothness penalty-based PWLS into the Karhunen-Loève (KL) domain for LDCT reconstruction [3]. With the development of compressed sensing (CS) theory [4, 5], it was proven that the sparsity regularized by the L1 norm exhibited superior performance. Yu *et al*. [6] and Sidky *et al*. [7, 8] first utilized the sparsity of the discrete gradient, which can be formulated as the total variation (TV) for LDCT reconstruction. To alleviate the piecewise smoothness in clinical practice caused by the inadequate assumption of the TV, many methods have been proposed. Niu *et al*. introduced a higher order gradient and proposed the total generalized variation (TGV) reconstruction model for sparse-view computed tomography (CT) [9]. By learning a redundant dictionary, Xu *et al*. [10] more efficiently explored the sparsity of LDCT signals and provided promising results. Other representative methods include nonlocal means filtering [11, 12], tight wavelet frames [13], low-rank techniques [14], transform learning [15,16], convolutional sparse coding [17], etc. These algorithms usually achieve satisfactory results by employing artful handcrafted regularization terms and carefully determining the model parameters. However, both steps are empirical and laborious, which make it difficult to generalize to different images from different scanning protocols or parts of the human body.

Inspired by the success of deep learning [18, 19] in many related fields, such as image processing and computer vision, learning-based methods have become the mainstream of medical imaging [20, 21]. These proposed methods can be roughly classified into two categories: 1) the image-to-image method and 2) the data-to-image method. The first belongs to the postprocessing method. These methods do not need to access the projection data, circumventing the encrypted data protocols imposed by the scanner manufacturers. A neural network is trained to map the LDCT image directly into the corresponding normal-dose CT (NDCT) image. The first work in this field is a three-layer convolutional neural network (CNN) proposed by Chen *et al*. [22]. However, due to the shallow network depth, the results indicate that some structures may be blurred. To recover more details, Kang *et al*. [23] performed a wavelet transform on the input LDCT image and fed the high frequency components into the U-Net [24]. By using skip connections, Chen *et al*. developed a residual encoder-decoder CNN (RED-CNN) for LDCT denoising [25]. Shan *et al*. compared a modularized CNN with typical iterative reconstruction methods from three well-known vendors and showed competitive performance for LDCT image reconstruction [26]. Han *et al*. [27] and Jin *et al*. [28] improved U-Net to suppress the artifacts caused by undersampling. To make the predicted images obey the same statistical distribution as NDCT, a generative adversarial network (GAN) was introduced to LDCT [29, 30], and a discriminator network was employed to implement this constraint. Shan *et al*. [31] extended the pretrained 2D network to a 3D model with the help of transfer learning. In [32], the authors introduced the attention mechanism in both the plane and depth channels for 3D networks. These methods can obtain good results with fast computational speed. However, since these network models usually adopt a pseudo-inversion (such as FBP) algorithm as the input and neglect the relationship with the original measured projection data, the data consistency cannot be guaranteed. Therefore, this kind of method is criticized for its robustness. The second kind of method can partially address this challenge. In [33], the authors proposed a network model called AUTOMAP to simulate the imaging procedure directly from the measured data to the images. This method fully demonstrates the power of deep learning, but due to the high-dimensional data, the scale of the network is usually large, which significantly hampers its practical application. To mitigate the impact of the data dimension, [34] and [35] independently introduced imaging physics into this model and efficiently reduced the model scale. Inspired by the idea of sparse coding, unrolling the specific numerical scheme into a network is another popular way to integrate the imaging models into the reconstruction network. Chen *et al*. unrolled the steepest gradient descent algorithm and proposed the learned experts' assessment-based reconstruction network (LEARN) for sparse-view CT [36]. Adler and Öktem generalized the primal-dual hybrid gradient (PDHG) algorithm by replacing both the primal and dual proximal operators with learned operators, which were implemented by a trained CNN [37]. Similarly, the authors in [38] substituted the projector in the projected gradient descent algorithm with a CNN and imposed the measurement consistency into the unrolled network. With the idea of plug-and-play [39], He *et al*. performed convolutions in the image domain as learned regularization terms and plugged the intermediate result back to the alternating direction method of multipliers (ADMM) framework [40]. Since this kind of method involves a projection to correct the intermediate image in each iteration block, the results usually have a higher reconstruction accuracy, which is of great clinical importance for medical diagnosis. However, as mentioned in [32], spatial convolution is a local operator only focused on adjacent pixels, ignoring the fact that CT image data are located on a low-dimensional manifold, which accommodates rich topological structure information [41, 42].

In this paper, to simultaneously extract the pixel-level and topological features of LDCT data, we propose a manifold and graph integrative convolutional (MAGIC) network that performs in both image and manifold spaces for

LDCT reconstruction. First, we unroll the steepest gradient descent algorithm into a neural network and use a CNN module to replace the handcrafted regularization terms. Then, to introduce the low-dimensional manifold features, overlapped patches with a small size are extracted from the image to form a patch set. This operation is based on a well-accepted assumption that the patch set is located on a low-dimensional smooth manifold referred to as a patch manifold [41-44]. Since spatial convolution cannot process such data, inspired by the success of a graph convolution [45], we construct a graph using the points sampled from the patch manifold, and a graph convolution is applied to extract the topological features from the graph. In addition, since it is difficult to obtain a large amount of paired low-dose and normal-dose data in clinical practice, our proposed method alleviates this drawback by introducing a projection loss, which enables our semi-supervised learning model.

The remainder of this paper is organized as follows. In the next section, the details of our proposed method are elaborated. In the third section, the experimental results are presented. The last section provides the discussion and conclusion.

## 2. Methodology
### A. LEARN network for CT reconstruction

Different numerical schemes can be unrolled into a neural network. For simplicity, the gradient decent-based LEARN model is chosen as the backbone of our proposed method. To make this paper self-contained, in this section, we briefly introduce LEARN [36]. A general model for regularized reconstruction is as follows:

$$\min_x \tfrac{1}{2}\|Ax - y\|_2^2 + \lambda R(x), \tag{1}$$

where $x \in \mathbf{R}^{M_2}$ denotes the vectorization of image $f \in \mathbf{R}^{m \times n}$ ($M_2 = m \times n$), $y \in \mathbf{R}^{M_1}$ represents the measured projection data, and $A \in \mathbf{R}^{M_1 \times M_2}$ is the system matrix, in which each element $a_{i,j}$ stands for the contribution to the $i$-th projection of the $j$-th pixel. $R(x)$ denotes the regularization term reflecting the prior knowledge of the image to reconstruct, and $\lambda$ is a weight to balance the measurement and regularization term. To incorporate the model into the deep learning technique framework, a generalized regularization term, referred to as the field of experts (FoE) [46], is introduced as

$$R(x) = \sum_{k=1}^{K} \psi_k(\varphi_k x), \tag{2}$$

where $\varphi_k$ and $\psi_k$ denote the regularization operator and potential function, respectively, both of which can be learned from the existing dataset. Then, we can obtain $x$ by solving the model,

$$\hat{x} = \arg\min_x \tfrac{1}{2}\|Ax - y\|_2^2 + \sum_{k=1}^{K} \lambda_k \psi_k(\varphi_k x). \tag{3}$$

A simple steepest gradient descent algorithm can be applied to Eq. (3):

$$x^{t+1} = x^t - \alpha \left[ (Ax^t - y) + \sum_{k=1}^{K} \lambda_k \varphi_k^* \psi_k'(\varphi_k x^t) \right], \tag{4}$$

where $\varphi^*$ represents the conjugate operator of $\varphi$, and $\alpha$ is the step size. The iteration-dependent form of Eq. (4) can be written as

$$x^{t+1} = x^t - \alpha^t(Ax^t - y) + \sum_{k=1}^{K} \lambda_k^t \varphi_k^{t*} \eta_k^t(\varphi_k^t x^t) \tag{5}$$

where $\eta(\cdot) = \psi'(\cdot)$. The last term in Eq. (5) performs spatial filtering, which can be generalized as a neural network, and in LEARN, a three-layer CNN module is utilized to substitute it as

$$x^t = x^t - \alpha^t(Ax^t - y) + \Phi(x^t), \tag{6}$$

in which

$$\Phi(x^t) = w_3^t * \sigma(w_2^t * \sigma(w_1^t * x^t)), \tag{7}$$

where $w$ is the trained kernel, $*$ denotes the convolution operator and $\sigma(\cdot)$ is the activation function. Once the iteration number is fixed, we can unroll Eq. (6) into a network with a determined number of layers. The initial reconstruction $x^0$, projection $y$, and system matrix $A$ are input into the network. Notably, Eq. (6) can be seen as a residual block, which is composed of three parts: a skip connection, a data fidelity layer and a spatial CNN module.

**B. Patch manifold and graph convolutional network**

To extract nonlocal topological features of the LDCT data, we sample from the patch manifold to study the low-dimensional manifold of the LDCT data. Considering an LDCT image with $m \times n$ pixels $f \in \mathbf{R}^{m \times n} = \{f(i,j) | 1 \le i \le m, 1 \le j \le n\}$, we extract a small rectangular patch $p_{ij}(f)$, which has pixel $f(i,j)$ as the top-left corner and a size of $s_1 \times s_2$. We fix the acquisition step size and collect the patches to obtain a set:

$$\mathcal{P}(f) = \{p_{ij}(f) | (i,j) \in \{1, 1+i_0, 1+2i_0, \dots, m\} \times \{1, 1+j_0, 1+2j_0, \dots, n\}\} \subset \mathbf{R}^d, d = s_1 \times s_2, \tag{8}$$

where $i_0 \in [1, s_1]$ and $j_0 \in [1, s_2]$ represent the acquisition step sizes, which guarantee that the collected patches overlap. $\mathcal{P}(f)$ can be seen as a point cloud sampled from a low-dimensional manifold $\mathcal{M}(f)$ embedded in $\mathbf{R}^d$, referred to as the patch manifold associated with $f$. The patch manifold is of a low dimension for different kinds of images [41-44].

Since the spatial convolution cannot handle the sampling points from the patch manifold, in this paper, a graph neural network is constructed with these points, and graph convolution is utilized to extract the features of the low-dimensional manifold, which has shown great potential in classification work [45, 47-49].

Once the patch set $\mathcal{P}(f)$ is obtained, we can construct a graph $\mathcal{G}(\mathcal{V}, \mathcal{E})$ with N nodes, each of which is the element of $\mathcal{P}(f)$. The adjacency matrix $W \in \mathbf{R}^{N \times N}$ of the graph can be calculated with a Gaussian function [41]:

$$W_{ij} = \exp\left(-\frac{\|v_i - v_j\|_2^2}{\sigma(\mathcal{V})^2}\right), \tag{9}$$

where $v_i, v_j \in \mathcal{V}$ are the two nodes in the graph and $\sigma(\mathcal{V})$ is the standard deviation of the nodes. The Euclidean distance is adopted to measure the distances between adjacent nodes, and we take the median as the estimation of the standard deviation for simplicity. The diagonal degree matrix $D$ is defined as $D_{ii} = \sum_j W_{ij}$.

Once the graph is constructed, the spectral graph convolution [47] can be used to study the graph:

$$g_\theta * a = U g_\theta U^T a, \tag{10}$$

where $a \in \mathbf{R}^N$ is the signal whose element is a scalar node, $g_\theta = diag(\theta \in \mathbf{R}^N)$ denotes the trained kernel, and $U$ is the matrix of eigenvectors of the normalized graph Laplacian $L = I - D^{-\frac{1}{2}} W D^{-\frac{1}{2}} = U \Lambda U^T$ in which $I$ represents the identity matrix. To reduce the computational complexity, Hammond *et al.* [48] proposed to approximate Eq. (10) with truncated Chebyshev polynomials:

$$g_\theta * a \approx \sum_{k=0}^{K} \theta_k T_k(\tilde{L}) a, \tag{11}$$

where $\tilde{L} = \frac{2}{\lambda_{max}} L - I$, in which $\lambda_{max}$ is the largest eigenvalue of $L$, and $T_k(\cdot)$ is the Chebyshev polynomial defined as

$$T_k(b) = \begin{cases} 2b T_{k-1}(b) - T_{k-2}(b), & k \ne 0 \text{ and } k \ne 1 \\ 1, & k = 0 \\ b, & k = 1 \end{cases}. \tag{12}$$

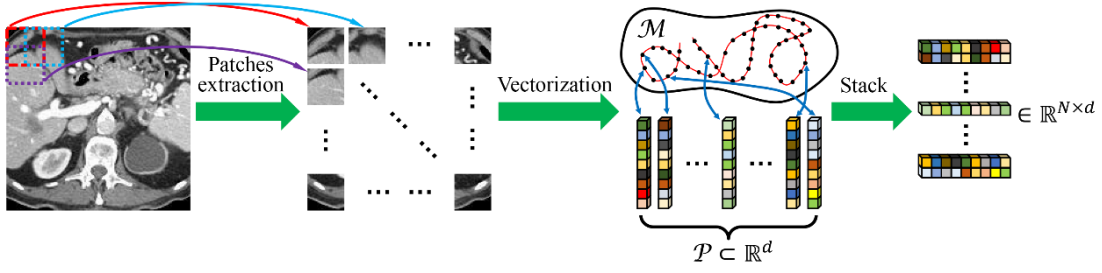

Fig. 1. Diagram of the linear transform $I$. Each vectorized patch corresponds to (blue arrows) a black point on the patch manifold. The patch set $\mathcal{P}$ (black points) has a trivial 2D parameterization (red curve) on the patch manifold $\mathcal{M}$.

Eq. (11) is the $K^{\text{th}}$-order polynomial, which means that the convolution on each node will involve the nodes that are at a maximum K steps away from the node. The spectral convolution is $K$-localized, and the involved nodes are $K^{\text{th}}$-order neighborhoods [49]. Let $K = 1$, $\lambda_{\max} \approx 2$ and $\theta = \theta_0 = -\theta_1$, we have

$$g_\theta * a \approx \theta_0 a + \theta_1 (L-I)a = \theta \left( I + D^{-\frac{1}{2}} W D^{-\frac{1}{2}} \right) a, \tag{13}$$

which is equal to the graph convolutional network (GCN) proposed by Kipf *et al.* [45]. With the renormalization trick proposed in [45], let $\widetilde{W} = I + W$ and $\widetilde{D}_{ii} = \sum_j \widetilde{W}_{ij}$, we have $I + D^{-\frac{1}{2}} W D^{-\frac{1}{2}} \to \widetilde{D}^{-\frac{1}{2}} \widetilde{W} \widetilde{D}^{-\frac{1}{2}}$.

For a signal $X \in \mathbf{R}^{N \times d}$, which has $N$ nodes with a $d$-dimensional feature vector, GCN can be formulated as [43]

$$Z = \widetilde{D}^{-\frac{1}{2}} \widetilde{W} \widetilde{D}^{-\frac{1}{2}} X \Theta. \tag{14}$$

where $\Theta \in \mathbf{R}^{d \times F}$ is the trained filter and $Z \in \mathbf{R}^{N \times F}$ is the convolved signal matrix.

### C. The Proposed MAGIC for LDCT reconstruction

In our method, we attempt to simultaneously extract the pixel-level and topological features by incorporating both spatial and graph convolutions. In Eq. (6), a three-layer CNN module is used to extract the local pixel-level features of $x^t$. To impose the nonlocal topological features from the low-dimensional manifold space, we modified Eq. (6) and added a GCN term. First, a patch set $\mathcal{P}(f^t)$ is built, and we construct a graph $\mathcal{G}^t(\mathcal{V}, \mathcal{E})$ with N nodes, each of which corresponds to a certain element of $\mathcal{P}(f^t)$. Then, the nodes are stacked to obtain the matrix signal $X^t \in \mathbf{R}^{N \times d}$, and two successive graph convolutions are applied on it. Our model is modified from Eq. (6) to:

$$x^{t+1} = x^t - \alpha^t (Ax^t - y) + \Phi(x^t) + \Psi(X^t), \quad X^t = I(f^t) \tag{15}$$

where $\Psi(X^t) = \widetilde{D}^{-\frac{1}{2}} \widetilde{W} \widetilde{D}^{-\frac{1}{2}} \sigma \left( \widetilde{D}^{-\frac{1}{2}} \widetilde{W} \widetilde{D}^{-\frac{1}{2}} X^t \Theta_1^t \right) \Theta_2^t$, $\Theta_1 \in \mathbf{R}^{d \times F}$ and $\Theta_2 \in \mathbf{R}^{F \times d}$ are the graph convolutional kernels and $I: \mathbf{R}^{m \times n} \to \mathbf{R}^{N \times d}$ is the linear transform to obtain $X^t$. Fig. 1 illustrates the main steps to obtain $X$ from $I$. In Fig. 1, the vectorized patches correspond to the points on the smooth manifold, and $\mathcal{P}$ has a trivial 2D parameterization $(i,j) \to p_{ij}$ on the patch manifold.

Notably, the computation of the adjacency matrix is time-consuming if we update it in each iteration. Based on this consideration, we divide the whole iteration procedure into two stages: coarse and fine stages. Fig. 2 shows the flowchart of our proposed unrolled iteration network MAGIC. In the coarse stage, $x^0$ (initial reconstruction with FBP) and projection data $y$ are fed into the network. Compared with LEARN, one parallel path, which performs graph convolution, is added into each iteration block. The adjacency matrix of coarse stage $W_C$ is calculated based on $x^0$ and kept fixed in each iteration block during the entire coarse stage. The graph transform in Fig. 2 equals the linear function: $I$, and the inverse graph transform denotes the inverse operator of $I$. In the coarse stage, the result of FBP usually suffers from heavy noise, which makes $W_C$ inaccurate. After the $t + 1$ iteration, once the noise of $x^{t+1}$ has been basically removed, the network enters the fine stage. We recalculate the adjacency matrix $W_F$ based on $x^{t+1}$ and leave it unchanged during the entire fine stage.

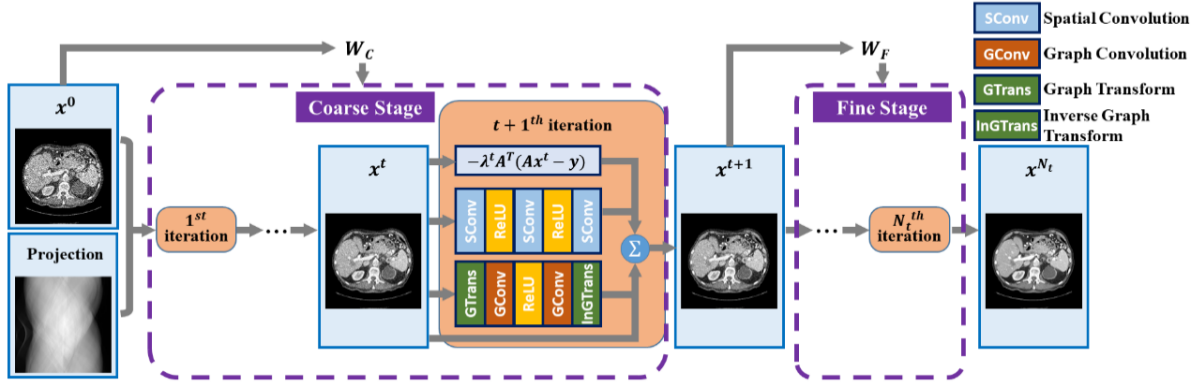

Fig. 2. Illustration of our proposed MAGIC.

For simplicity, the mean square error (MSE) is adopted as the loss function:

$$L_{MSE} = \frac{1}{N_s}\sum_{i=1}^{N_s}\|x_i - \hat{x}_i\|_2^2, \quad (16)$$

where $x_i$ is the predicted reconstruction result and $\hat{x}_i$ is the corresponding label. $N_s$ is the total number of samples. In addition, we apply our proposed MAGIC to only part of the labeled samples. The projection loss is proposed as:

$$L_{Pro} = \frac{1}{N_s}\sum_{i=1}^{N_s}\|Ax_i - y_i\|_2^2, \quad (17)$$

where $y_i$ is the corresponding measured projection data. In the case of semi-supervised learning, the loss function can be formulated as:

$$L = \frac{1}{|S_1|}\sum_{i\in S_1}\|x_i - \hat{x}_i\|_2^2 + \frac{1}{|S_2|}\sum_{i\in S_2}\|Ax_i - y_i\|_2^2 \quad (18)$$

where $S_1$ and $S_2$ are the sets of labeled and unlabeled samples, respectively. $|S_1|$ and $|S_2|$ denote the numbers of elements in $S_1$ and $S_2$, respectively, and $|S_1| + |S_2| = N_s$. While dealing with the unlabeled data in the training set, the projection loss can be leveraged to avoid overfitting. Importantly, a more advanced loss function or network architecture, such as perceptual loss or GAN, may further boost the model performance, but in this paper, we only focus on exploring the power of integrating both spatial and graph convolutions in the image and manifold spaces and making a fair comparison with other existing models.

## 3. Experiments and results

To evaluate the performance of our MAGIC, the dataset "*the 2016 NIH-AAPM-Mayo Clinic Low-Dose CT Grand Challenge*" [50] was used in our experiments. The dataset has 5936 full-dose CT images from 10 patients. In our experiments, 400 images were randomly selected from 8 patients as the training set, and 100 images were chosen from the remaining 2 patients as the test set. The size of the image was $256 \times 256$. The projection data were simulated with the distance-driven method [51, 52]. The distances of the X-ray source and detector to the rotation center were both 25 cm. The physical height and width of a pixel were both 0.6641 mm. The detector had 512 elements, each of which had a length of 0.72 mm. On average, 1024 projection views were sampled in the 360 degree range. To simulate a realistic clinical environment, Poisson noise and electronic noise were added into the measured projection data as [9]:

$$y = \ln\frac{I_0}{\text{Poisson}(I_0\exp(-\hat{y})) + \text{Normal}(0, \sigma_e^2)}, \quad (19)$$

where $I_0$ is the number of photons before the X-rays penetrate the object, $\sigma_e^2$ is the variance of electronic noise caused by the equipment measurement error, and $\hat{y}$ represents the noise-free projection. In our experiments, the X-ray intensity of a normal dose was set to $I_0 = 10^6$ according to [9]. Three different dose levels were simulated as low-dose cases, including 10%, 5% and 2.5%, i.e., $I_0 = 10^5, 5 \times 10^4,$ and $2.5 \times 10^4$, respectively. In all the experiments, we fixed the electronic noise variance at $\sigma_e^2 = 10$.

The size of the spatial convolution kernels was set to $3 \times 3$. When sampling from the patch manifold, the extracted patch size was set to $6 \times 6$, and the acquisition step size was $i_0 = j_0 = 2$. While calculating the adjacency matrix, 8 nearest neighbors of each node were included to make the adjacency matrix sparse and reduce the computational complexity. The sizes of the graph convolution parameters $\Theta_1$ and $\Theta_2$ were $36 \times 64$ and $64 \times 36$, respectively. The number of iterative blocks was fixed to 50. The coarse and find stages had 25 blocks. In the semi-supervised learning experiments, only 10% of the training data, which means only 40 images have labels. During the semi-supervised training, inputs of the network were randomly selected from both labeled and unlabeled data. In a single batch, while both labeled and unlabeled data exist, the losses were separately calculated and then summed.

Meanwhile, to verify the clinical feasibility of our proposed method, real data were also tested using the network trained with simulated data. In addition, we conducted extensive experiments to verify the robustness of our proposed model for different cases. The experiments were performed in Python 3.6 with the PyTorch library on a PC (Intel Core i5 8400 CPU, 16 GB RAM and GTX 1080Ti GPU). Our codes for this work are available on https://github.com/xwj01/MAGIC.

Four state-of-the-art methods were involved for comparison, including TGV [9], RED-CNN [25], learned primal-dual (LPD) [37] and LEARN [36]. All the implementations of these methods were provided by the original authors. TGV is a regularization-based method that utilizes the sparsity of the high-order gradient of images. RED-CNN is a supervised learning-based postprocessing method for LDCT image restoration. LPD and LEARN are both unrolling iteration network methods that adopt spatial convolution for image filtering. The semi-supervised learning version of MAGIC is referred to as MAGIC-Semi. The training epochs of all the learning-based methods were fixed to 100. The peak signal-to-noise ratio (PSNR) and structural similarity index measure (SSIM) were employed to quantitatively evaluate the performance of different methods.

**A. Validation with the simulated data**

Fig. 3 shows the results of an abdominal image reconstructed by different methods with a 10% dose. The FBP results suffer from severe noise. The TGV method removes most of the noise while preserving the details to a certain degree. The learning-based methods can also effectively suppress the noise, and most of the detailed information is well maintained in these results. Two possible metastases, which are indicated by the blue arrows, are apparent in all the results of Fig. 3. However, some oversmoothed effects can also be observed in the results of the learning-based methods. Without the help of the measured data, the detailed distortion in the result of the RED-CNN is obvious, especially for the contrast-enhanced vessels in the liver. Although LEARN and LPD were deduced from different numerical schemes, after adequate training, they achieved similar performance. The proposed MAGIC and MAGIC-Semi obtained the best visual result and preserved most details. In the region indicated by the red arrow, the vascular structures in our results are more complete and have a higher contrast than the other methods. It can also be seen that our proposed models achieved noticeable improvements in terms of both the PSNR and SSIM.

To better visualize the performance of different methods, we magnify the region indicated by the red rectangle in Fig. 3 (a). The enlarged parts of the different methods are presented in the bottom right corner. The shapes of the lesions in the reconstructed results with different methods are basically distinguishable. However, there are still some visible blocky artifacts in TGV result. Two purple arrows indicate two minute vessels, and only TGV and our methods recovered them well. All the other methods blurred these details to varying degrees. In the area indicated by the yellow arrow, TGV result produced piecewise smooth result. Although the result of MAGIC-Semi has more mottle-like noise than that of MAGIC, the visual effect is more similar to the ground truth data, which is more compatible with the doctors' reading habits.

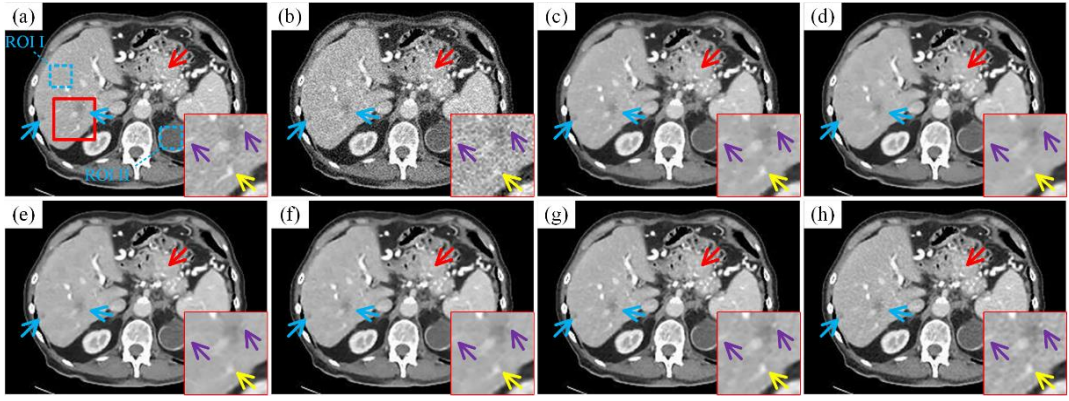

Fig. 3. Abdominal reconstruction with 10% dose data by different methods. (a) Ground truth, (b) FBP (25.02/0.7084), (c) TGV (30.84/0.8925), (d) RED-CNN (31.20/0.8955), (e) LPD (31.38/0.9040), (f) LEARN (31.80/0.9078), (g) MAGIC (**34.00**/0.9356) and (h) MAGIC-Semi (33.55/**0.9360**). The display window is [-160, 240] HU.

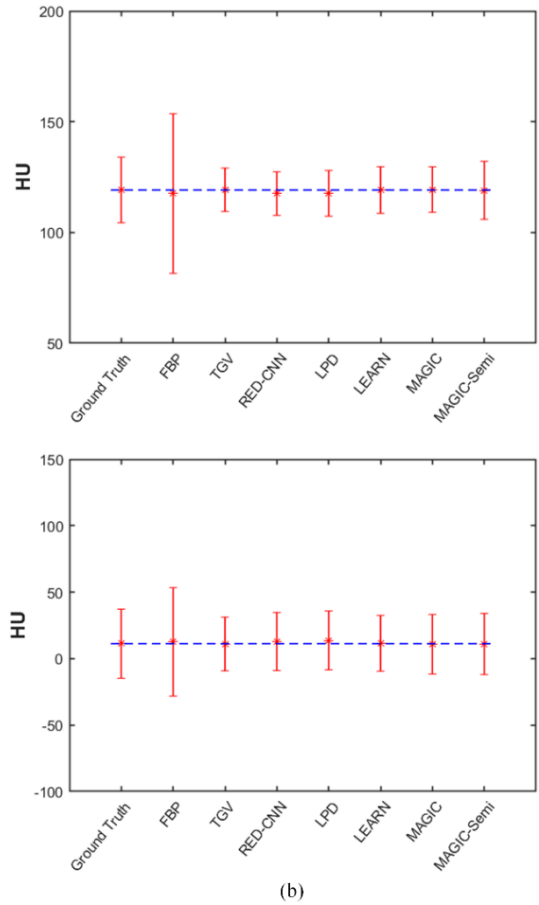

Fig. 4 Means and standard deviations for (a) ROI I and (b) ROI II obtained with different methods.

To quantitatively study the reconstruction error in local regions, two homogeneous regions of interest (ROIs) indicated by two blue dotted boxes in Fig. 3 (a) were selected to calculate the means and standard deviations (SDs), and the results are depicted in Fig. 4. It can be seen that FBP has obvious biases with regard to both the means and SDs. Since RED-CNN is a postprocessing algorithm and the input is the result of FBP, it has a similar means to FBP but a smaller SD. LPD also shows similar results to those of RED-CNN. The remaining four methods obtained an accurate means to the ground truth. Specifically, the result of MAGIC-Semi has the closest SD to that of the ground truth, which agrees with the observation in Fig. 3 that MAGIC-Semi can partially maintain the mottle-like texture in the ground truth.

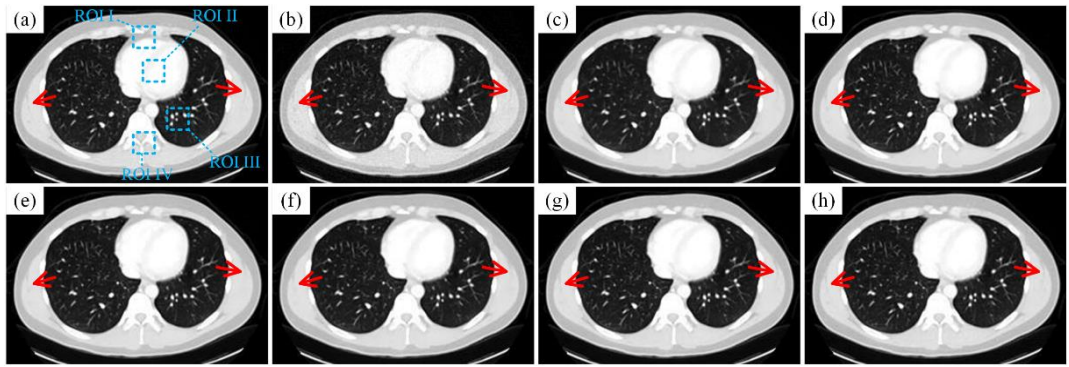

Fig. 5. Thoracic reconstruction with 10% dose data by various methods. (a) Ground truth, (b) FBP (27.93/0.7762), (c) TGV (31.49/0.9299), (d) RED-CNN (32.98/0.9423), (e) LPD (33.24/0.9521), (f) LEARN (33.74/0.9517), (g) MAGIC (**36.26/0.9696**) and (h) MAGIC-Semi (35.58/0.9692). The display window is [-1000, 200] HU.

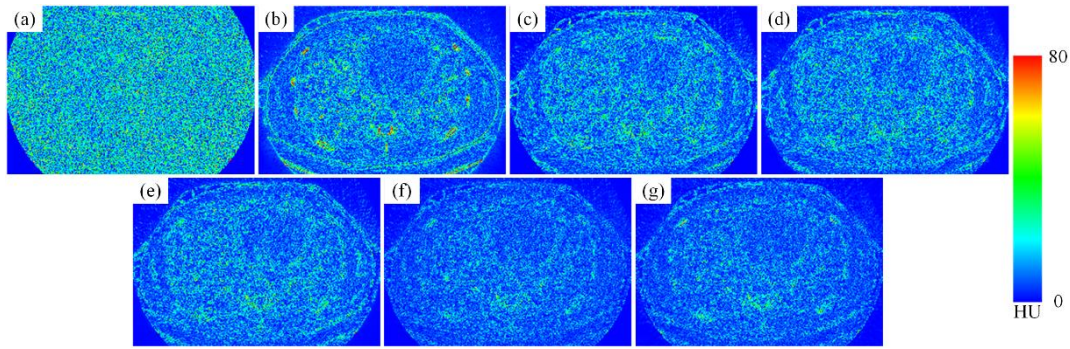

Fig. 6. Absolute difference images associated to the ground truth (a) FBP, (b) TGV, (c) RED-CNN, (d) LPD, (e) LEARN, (f) MAGIC and (g) MAGIC-Semi.

Fig. 5 demonstrates the reconstructions of a thoracic slice using different methods, and the results are displayed in the lung window. All the methods can efficiently eliminate the noise in the current display setting. Two red arrows indicate that two edges can visually differentiate the performance of different methods. Only MAGIC preserved these structures well, and other methods smoothed them to varying degrees. To better visualize the denoising performance of different methods, we show the absolute difference images associated with the ground truth in Fig. 6. It is clear that our proposed methods yielded the smallest difference from the ground truth, eliminating most noise and maintaining more details. Consistent with the visual results, our proposed methods have the best PSNR and SSIM, and the improvement is significant.

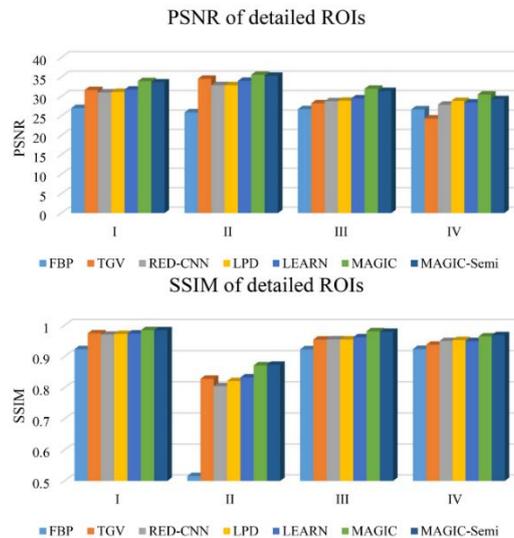

Fig. 7. Quantitative results of different methods over the ROIs indicated in Fig. 5 (a).

Four typical ROIs, which are indicated by the blue boxes (ROI I-IV) in Fig. 6 (a), were chosen to calculate the local PSNR and SSIM. The results are given in Fig. 7. Generally, our proposed methods obtained better scores in terms of both metrics in all four ROIs. MAGIC had slightly better PSNR values than MAGIC-Semi, and the SSIM values of both models were similar.

Fig. 8 demonstrates the training loss curves of different learning-based methods. It can be seen that 100 epochs are enough for all the methods to converge stably, and our network converges faster than all the other methods with the lowest loss.

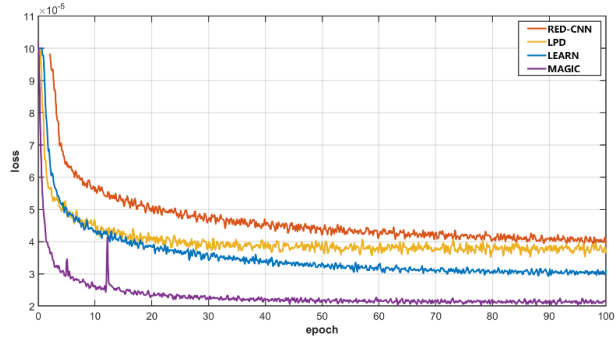

Fig. 8. The loss curves of different methods with data from the 10% dose.

**B. Validation with real data**

To study the clinical potential of our proposed method, the real data were tested using networks trained with simulated data. The real data were obtained by circular cone-beam scanning of a sheep lung with a Siemens Somatom Sensation 64-slice CT. NDCT data were acquired at 100 kV and 150 mAs, and LDCT data were obtained after the injection of a contrast agent at 80 kV and 17 mAs (due to the unavoidable registration problem, there are some inconsistent structures between Fig. 10 (a) and (b)). The cone-beam sinograms were rearranged into fan-beam type. The distance between the X-ray source and rotation center was 57 cm. A total of 1160 projection views were uniformly sampled in the 360 degree range. The detector had 672 elements, each of which covered an angle of 0.0014 rad, and the radius of field-of-view (FOV) was 25.05 cm. In this experiment, the scanned area was divided into a $256 \times 256$ grid, covering an area of $29.09 \times 29.09$ cm$^2$. Fig. 9 shows the reconstruction using different methods. The networks of learning-based methods were trained with simulated data of 10% dose as in the previous subsection. Notably, the geometry of the training data is not the same as that of scanning the sheep lung, which means our model can be easily extended to different datasets with different scanning geometries. It can be seen that the noise displayed in Fig. 9 (b) was basically removed using different methods. RED-CNN, LPD and LEARN still suffer from noise, as indicated by the red arrow. In this respect, our proposed methods show better robustness. In the region indicated by the blue arrow, our proposed MAGIC also has the clearest edges. To better observe the details of reconstructions, the area indicated by the red rectangle in Fig. 9 (a) is magnified and presented in the bottom left corner. The vessels indicated by the red arrows of RED-CNN, LPD and LEARN are blurred, and TGV and our proposed methods recovered them well.

**C. Study of different dose levels**

To evaluate the robustness of MAGIC, two more datasets simulated with 5% and 2.5% doses were used. The results of a femur case with 5% dose data and a pelvis case with 2.5% dose data are shown in Figs. 10 and 11, respectively. Due to the significantly reduced radiation dose, the FBP results are severely degraded. In Fig. 10, the noise can be almost removed using different methods. In the regions indicated by the red arrows, the results of RED-CNN, LPD and LEARN can hardly maintain the structures. TGV and our methods better preserved these details, but TGV left some noise in the middle of the image. In Fig. 11, as the radiation dose was seriously reduced, the blurring effect became more obvious. The structure indicated by the red arrow is missing in the results of RED-CNN, LPD and

LEARN but can be recognized in the results of TGV and MAGIC. However, TGV cannot recover the structure indicated by the blue arrow.

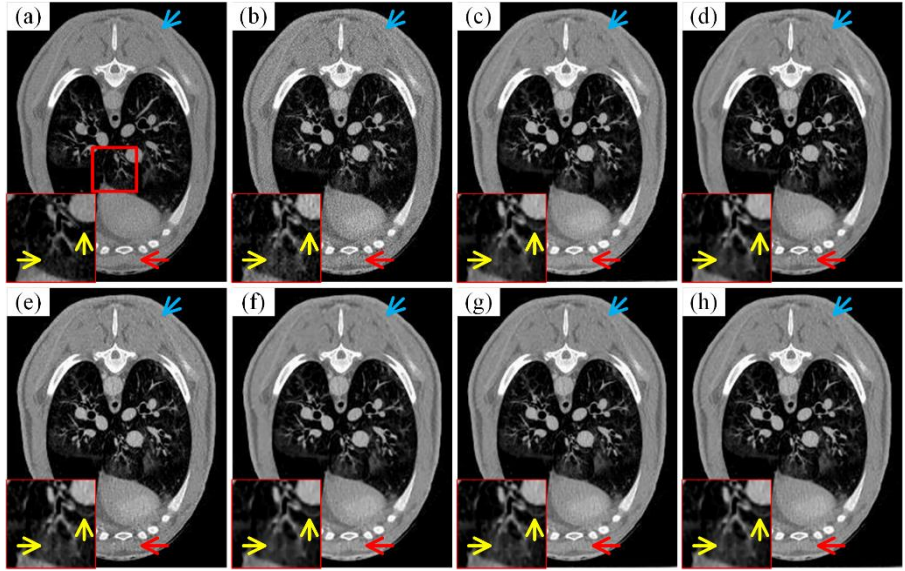

Fig. 9. Reconstructions of real data with different methods. (a) FBP of NDCT data, (b) FBP of LDCT data, (c) TGV, (d) RED-CNN, (e) LPD, (f) LEARN, (g) MAGIC and (h) MAGIC-Semi. The display window is [-555 575] HU.

The statistical quantitative results of the whole testing set using different learning-based methods are shown in Table I, which gives the means and SDs of PSNR and SSIM. It is clear that our methods obtained higher scores than all the other methods. It is worth mentioning that the results of MAGIC-Semi have lower scores than those of MAGIC, which is not exactly coherent with the visual results. This result is probably because the L2 norm loss will lead to higher PSNR and SSIM values but smoother images [30].

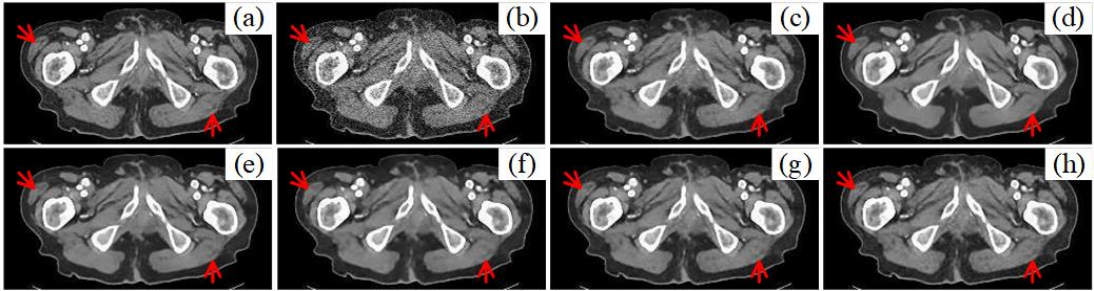

Fig. 10. Femoral reconstruction with 5% dose data by different methods. (a) Ground truth, (b) FBP (24.25/0.6142), (c) TGV (31.38/0.9229), (d) RED-CNN (31.91/0.9227), (e) LPD (32.04/0.9328), (f) LEARN (32.72/0.9404), (g) MAGIC (**34.82/0.9562**) and (h) MAGIC-Semi (34.40/0.9545). The display window is [-160, 240] HU.

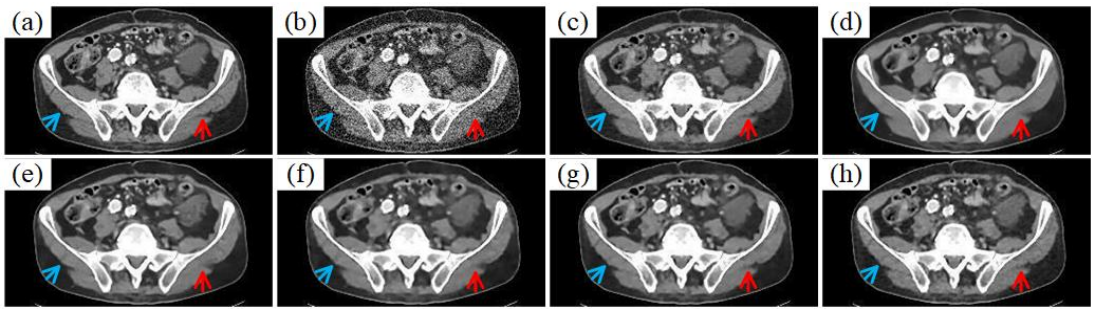

Fig. 11. Pelvic reconstruction with 2.5% dose data by different methods. (a) Ground truth, (b) FBP (21.12/0.5598), (c) TGV (29.51/0.8908), (d) RED-CNN (29.49/0.9109), (e) LPD (29.76/0.9087), (f) LEARN (30.02/0.9233), (g) MAGIC (**32.52/0.9488**) and (h) MAGIC-Semi (32.04/0.9447). The display window is [-160, 240] HU.

Table I Quantitative results (Mean±SD) using different methods. The best scores are marked in red, and the second best scores are marked in blue.

| dose | 10% | | 5% | | 2.5% | |
|---|---|---|---|---|---|---|
| | PSNR | SSIM | PSNR | SSIM | PSNR | SSIM |
| FBP | 26.35±0.68 | 0.6969±0.0321 | 23.56±0.74 | 0.6160±0.0363 | 20.67±0.77 | 0.5381±0.0374 |
| TGV | 31.91±0.50 | 0.9210±0.0097 | 31.12±0.52 | 0.9103±0.0119 | 29.41±0.62 | 0.8721±0.0207 |
| RED-CNN | 32.89±0.58 | 0.9251±0.0124 | 31.57±0.60 | 0.9123±0.0144 | 29.93±0.63 | 0.8911±0.0169 |
| LPD | 33.12±0.59 | 0.9356±0.0117 | 31.59±0.61 | 0.9194±0.0140 | 30.05±0.62 | 0.8981±0.0165 |
| LEARN | 33.51±0.60 | 0.9363±0.0112 | 32.18±0.61 | 0.9299±0.0116 | 30.38±0.61 | 0.9090±0.0142 |
| MAGIC | **35.89±0.66** | **0.9587±0.0092** | **34.18±0.64** | **0.9460±0.0107** | **32.72±0.64** | **0.9335±0.0120** |
| MAGIC-Semi | **35.18±0.59** | **0.9548±0.0092** | **33.70±0.59** | **0.9425±0.0111** | **32.16±0.57** | **0.9275±0.0133** |

**D. Proportion of the labeled training data**

Fig. 12 shows the impact of different proportions of labeled training data on the performance. The proportion of labeled data is from zero (unsupervised learning) to 50% (200 slices). It can be seen that there is a sharp rise from zero to 0.25% (only one slice). When the proportion of labeled data reaches 10% (40 slices), the performance improves slowly, which shows that our method can use only a small number of labeled data to achieve satisfactory results, which is quite meaningful for clinical practice.

Meanwhile, we conducted experiments to verify the robustness of our proposed method to different dose levels and different training samples in the supplemental materials, in which we also discuss the impacts of hyperparameters in our model.

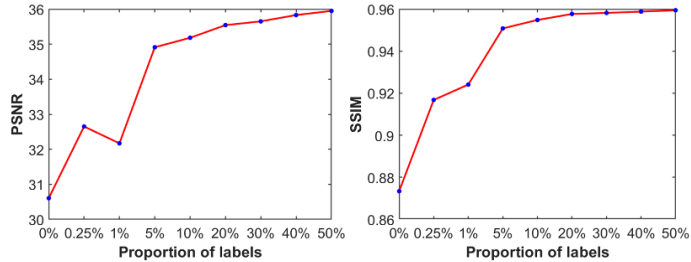

Fig. 12. Results of the networks trained with different proportions of labeled data.

## 4. Conclusion and discussion

In this paper, we propose a novel manifold and graph integrative convolutional network for LDCT reconstruction. This method not only uses spatial convolution to extract local pixel-level features in image space but also utilizes graph convolution to analyze the nonlocal topological features in manifold space. Compared with other methods, our method can capture the self-similarity during local pixels and nonlocal patches simultaneously. We conducted extensive experiments to evaluate the performance of our methods. Four state-of-the-art reconstruction methods, including TGV, RED-CNN, LPD and LEARN, were used for comparison. The results prove that our method outperforms other methods in both visual and quantitative aspects. In addition, our method is suitable for semi-supervised learning. In the case of only 10% of the data labeled, the results of our method have visually exceeded supervised learning-based methods. The success of MAGIC-Semi is of great significance for clinical applications, since paired low-dose and normal-dose data are quite difficult to obtain.

Our method has a limitation in that calculating the adjacency matrix requires the Euclidean distances between each two nodes to find the neighborhoods, which is time-consuming. We implemented this part with CUDA to accelerate the computation. Since we divide the network into coarse and fine stages, we only need to calculate the adjacency matrix twice during the training process. We must admit that the adjacency matrix used in each block is not the most accurate one. Fortunately, the neural network can fix this by adaptively updating the parameters. Table II shows the time cost of the different methods. Due to the computation of the adjacency matrix and the implementation of graph convolution, our method is more time-consuming than other learning-based methods.

Table II Time cost of different methods for training and testing

|  | TGV | RED-CNN | LPD | LEARN | MAGIC |
|---|---|---|---|---|---|
| Train | - | 1.7 h | 17.6 h | 20.8 h | 32.9 h |
| Test | 15.7 min | 2.6 ms | 1.8 s | 4.8 s | 5.3 s |

There are some potential issues we can consider to further improve our model. First, since the projection is contaminated, the data consistency layer may introduce error. Simultaneously, filtering both image and projection data may be a possible solution. Second, when our model processes data with different geometries as the training data, the performance cannot be guaranteed. In future work, we will focus on network architecture design and optimization to solve these limitations.

2013.

# Appendix

## A. Robustness analysis

1) Unpaired dose levels

To further study the robustness of our method, three models were trained with specific dose levels, including 10%, 5% and 2.5%, and these models were tested on datasets with various dose levels. Fig. 13 shows an abdominal case with different doses, and the reconstructions of MAGIC and MAGIC-Semi are shown in Figs. 14 and 15, respectively. It can be seen that the noise cannot be completely eliminated when using a network trained with higher-dose data to reconstruct lower-dose data. Conversely, the results become slightly blurred when using a network trained with lower-dose data to reconstruct higher-dose data. Notably, there are several metastases indicated by the red dotted circles, which are severely degraded by the noise. In all the results predicted by the networks with different training strategies, these metastases can be well recognized and have clear boundaries. A simple conclusion can be reached that although our proposed model cannot be generalized to an arbitrary dose level, it demonstrates robustness to a certain degree.

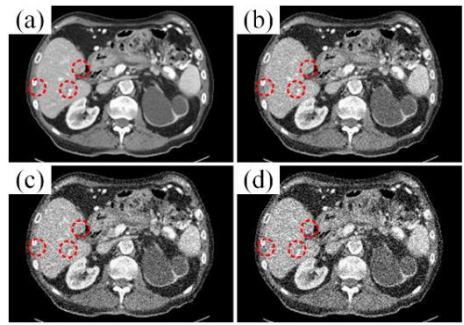

Fig. 13. Abdominal case of different doses. (a) Ground truth, (b) FBP of 10% dose data, (c) FBP of 5% dose data and (d) FBP of 2.5% dose data.

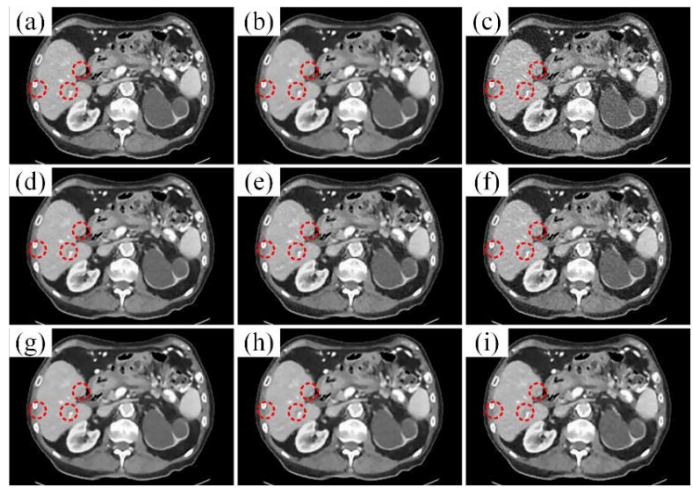

Fig. 14. The reconstructions with MAGIC trained with different dose data. From top to bottom, the results are obtained with the networks trained with 10%, 5% and 2.5% dose data. From left to right, the images are reconstructed by 10%, 5% and 2.5% dose data, respectively.

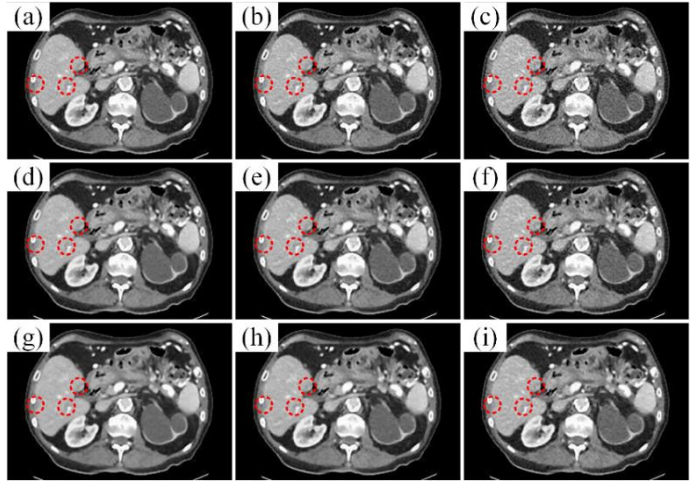

Fig. 15. The reconstructions with MAGIC-Semi trained with different dose data. From top to bottom, the results are obtained with the networks trained with 10%, 5% and 2.5% dose data. From left to right, the images are reconstructed by 10%, 5% and 2.5% dose data, respectively.

2) Different training samples

    We conducted another experiment to validate the robustness of our proposed model by training with natural images. We trained the network with a public dataset of natural images, DIV2K [53]. Four hundred images were randomly selected as the training set, and the same geometry was adopted as in Section 3 to obtain the projection data. Then, we tested the network with the 10% dose CT data. Fig. 16 shows the results. It can be seen that the network trained with natural images can eliminate the noise. The metastasis indicated by the dotted circles is well preserved. In particular, the metastasis indicated by the blue circle, which is difficult to identify, has clear boundaries in the results of both MAGIC and MAGIC-Semi trained with natural images. This finding can be treated as evidence that our method is robust to different training data.

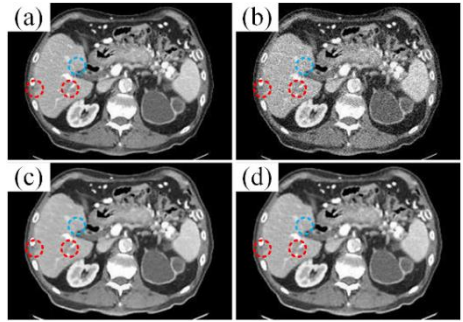

Fig. 16. Abdominal case of 10% dose data reconstructed with the networks trained with natural images. (a) Ground truth, (b) FBP, (c) MAGIC trained with DIV2K and (d) MAGIC-Semi trained with DIV2K.

**A. Hyperparameter study**

    In our method, there are several hyperparameters, including the number of iterative blocks $N_t$, the patch sizes of sampling the patch manifold $s_1$ and $s_2$, the size of the graph convolutional kernels $F$ and the proportion of labeled data for MAGIC-Semi. This subsection will discuss the impacts on the performance of these parameters.

1) Number of iterative blocks $N_t$

Generally, a deeper network will have better performance, but its training will be more difficult and require more memory. A proper depth is important to balance the tradeoff between the performance and other issues. Fig. 17 shows the quantitative results with different numbers of iterative blocks. It can be seen that the performance of the network improves rapidly as the network becomes deeper. When the number of iteration blocks reaches 50, the performance improvement slows down. Considering the performance and memory consumption, 50 is an appropriate number for the iteration blocks.

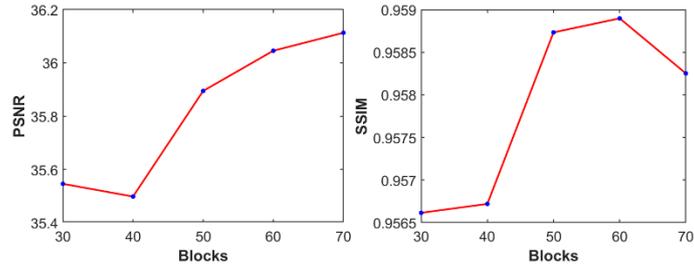

Fig. 17. Results of the networks with different numbers of iteration blocks.

2) Patch sizes $s_1$ and $s_2$

In our method, we extract square patches from the patch manifold, i.e., $s_1 = s_2$. If the patch size is too small, then the feature vector is insufficient to represent the characteristics of the LDCT data. If the patch size is too large, then it will boost the computational cost. The quantitative results with different patch sizes are shown in Table III. Based on this study, we set the patch size to 6 in our experiments.

Table III Quantitative results of different patch sizes

| Patch Size | PSNR | SSIM |
| --- | --- | --- |
| 4 | 35.66 | 0.9577 |
| 5 | 35.69 | 0.9576 |
| 6 | **35.89** | **0.9587** |
| 7 | 35.64 | 0.9579 |
| 8 | 35.67 | 0.9571 |
| 9 | 35.76 | 0.9581 |
| 10 | 35.80 | 0.9575 |

3) Size of graph convolutional kernels $F$

Table IV shows the performance of the network with different sizes of graph convolutional kernels. It can be noticed that the impact of $F$ is not very significant, so considering the visual effect, computational complexity, and memory issues, we set the size of graph convolutional kernels $F = 64$.

Table IV Quantitative results of different kernel sizes

| Kernel Size | PSNR | SSIM |
| --- | --- | --- |
| 32 | 35.86 | 0.9568 |
| 64 | 35.89 | 0.9587 |
| 96 | **35.97** | 0.9587 |
| 128 | 35.96 | **0.9588** |